\documentclass[a4paper,12pt]{article}
\usepackage{amssymb}
\usepackage{latexsym}

\usepackage{epsfig}
\usepackage{graphicx}

\usepackage{amsmath,amsthm,amsfonts,amssymb}

\newcommand{\bi}{\bibitem}
\newcommand{\be}{\begin{eqnarray}}
\newcommand{\ee}{\end{eqnarray}}

\newcommand{\BH}    {{\rm BH}}
\newcommand{\weak}  {{\rm W}}

\newcommand{\CP}    {{\rm CP}}

\newcommand{\SM}    {{\rm SM}}

\newcommand{\DW}    {{\rm DW}}
\newcommand{\sph}   {{\rm sph}}

\newcommand{\TBH}{T_{{\rm BH}}}
\newcommand{\mBH}{m_{{\rm BH}}}

\begin{document}

\begin{titlepage}
\begin{centering}
\vspace{.8in}
{\large {\bf Efficient electroweak baryogenesis by black holes}}
\\

\vspace{.5in}
{\bf Georgios Aliferis\footnote{aliferis@auth.gr},
\bf Georgios Kofinas\footnote{gkofinas@aegean.gr},
\bf Vasilios Zarikas\footnote{vzarikas@teilam.gr}}

\vspace{0.3in}
$^{1}$ Department of Physics\\
Aristotle University of Thessaloniki, 54124 Thessaloniki, Greece\\
\vspace{0.3cm}
$^{2}$ Research Group of Geometry, Dynamical Systems and Cosmology\\
Department of Information and Communication Systems Engineering\\
University of the Aegean, Karlovassi 83200, Samos, Greece\\
\vspace{0.3cm}
$^{3}$ Department of Electrical Engineering, Theory Division\\
ATEI of Central Greece, 35100 Lamia, Greece\\
\end{centering}

\vspace{1in}




\begin{abstract}

 A novel cosmological scenario, capable to generate the observed baryon number at the electroweak scale for
 very small CP violating angles, is presented. The proposed mechanism can be applied in conventional FRW
 cosmology, but becomes extremely efficient due to accretion in the context of early cosmic expansion with high
 energy modifications.
 Assuming that our universe is a Randall-Sundrum brane, baryon asymmetry can easily be produced by Hawking radiation of
 very small primordial black holes. The Hawking radiation reheats a spherical region around every black hole to
 a high temperature and the electroweak symmetry is restored there. A domain wall is formed
 separating the region with the symmetric vacuum from the asymmetric region where electroweak baryogenesis takes place.
 First order phase transition is not needed. The black holes's lifetime is prolonged due to accretion, resulting to
 strong efficiency of the baryon producing
 mechanism. The allowed by the mechanism black hole mass range includes masses that are energetically
 favoured to be produced from interactions around the higher dimensional Planck scale.

\end{abstract}

\end{titlepage}

\newpage

\baselineskip=18pt

\section{Introduction}

Baryogenesis is a key question of cosmology. Various interesting models of baryogenesis have been proposed during the
last thirty years (see reviews \cite{review1}-\cite{review6}).
An important piece of knowledge extracted from all this research is the realization that it has proved quite difficult
to construct a simple model capable to generate the observed amount of baryon asymmetry.

The standard mechanism of baryogenesis demands the concurrent satisfaction of three
physical requirements, as it was first explained by Sakharov~\cite{sakharov}:\\
1. Baryon non-conserving processes. This can be achieved either at the grand unification scale \cite{gut-B}, or even
at the electroweak energy scale \cite{ew-B}.\\
2. C and CP-violation which has already been observed in experiment. \\
3. Out of equilibrium conditions. This can be realized in an expanding universe in interactions evolving very
massive particles, or in a first order
phase transition, or due to thermal domain walls around black holes as proposed in the present work.

Baryogenesis by heavy particle decays was first proposed in \cite{sakharov}, \cite{kuzmin} (see also \cite{ad-bs},
\cite{ad-yz}). Due to heavy particle decays in an expanding cosmology in the presence of C and CP violation, baryon
asymmetry is produced. Usually this mechanism can be realized in
grand unification models with the heavy particle being a gauge boson of grand unification.
Electroweak baryogenesis is another scenario attracted a lot of study \cite{krs}. It is natural to expect to
generate baryon asymmetry at this energy scale, otherwise rapid sphaleron processes at $100$ GeV will destroy any
baryon asymmetry produced earlier.
The standard model incorporates non-conservation of baryons (through the chiral anomaly) \cite{ew-B},
as well as deviation from thermal equilibrium if the phase transition is
first order. Unfortunately, the recently discovered heavy Higgs boson turns the transition to a second order one.
Another disadvantage of the standard model is its very small CP violating phases \cite{ad-varenna}.
The present work refers to the electroweak energy scale, but solves both problems as it will be shortly explained.
There are of course other possibilities too to have baryogenesis at the electroweak scale, like TeV scale gravity
\cite{TeV-gravity}, \cite{TeV-grav-EW}.

Another way to produce baryon asymmetry is baryo-through-lepto-genesis \cite{fuku-yana}.
At high energies, large as $10^{10}$ GeV, lepton asymmetry is produced from heavy Majorana fermion decays, and
subsequently, this lepton asymmetry leads to baryon asymmetry by the equilibrium electroweak processes which break $(B+L)$
symmetry \cite{lgns}. A different mechanism is the Affleck-Dine baryogenesis \cite{affleck}.
In supersymmetric models scalar superpartners of baryons or leptons can acquire a large baryonic charge after
inflation. Subsequent $B$-conserving decay of these fields transform baryon asymmetry into that in the quark sector.
This mechanism, contrary to all others, leads to quite high value of baryon-to-photon ratio of order one, and special
effort is needed to generate the observed amount. The proposed mechanism in the current work is also able to generate
large baryon asymmetries which, however, can be controlled to have the correct value.
Large amounts of baryon asymmetry can also be generated in the so called Spontaneous baryogenesis \cite{spont-BS},
in which a spontaneously broken global ${U(1)}$ symmetry associated with baryonic number is assumed to exist.
Unfortunately, these models are associated to large isocurvature density perturbations at large scales which are
forbidden by cosmic microwave background.

Another interesting class of models concerns baryogenesis through evaporation of primordial black
holes (PBHs) \cite{zeld-bh}. The particle propagation of the Hawking radiation \cite{Hawking} in the gravitational field
of the BH distorts the thermal equilibrium and makes possible the creation of a net excess of particles over antiparticles,
assuming baryon non conserving heavy particle decays. The scenario presented here assumes the existence of primordial
black holes emitting thermal radiation. However, the mechanism for the generation of baryon asymmetry is totally different.
There are finally some more exotic scenarios, for example mechanisms based on space separation of $ B$ and $\bar B $
(these scenarios allow for baryonic charge conservation and globally baryo-symmetric universe) \cite{omnes}-\cite{ez}.
Successful baryogenesis could be realized even without the three Sakharov conditions \cite{ad-bs}, although these
mechanisms are somewhat more technically complicated.

Primordial black holes can be very small black holes created at the first moments of the universe
\cite{carr}. The first PBHs baryogenesis models were based on grand unified theories (GUT) \cite{GUT}. GUT processes can
truly produce baryon number, but this is subject to sphaleron wash out \cite{sphaleron}. The electroweak baryogenesis
scenario proposed by Cohen, Kaplan and Nelson (CKN model) \cite{CKN} addresses this problem applying the sphaleron
process to produce baryon number. Nagatani \cite{Nagatani}, in order to overcome the well known problems of
electroweak baryogenesis, proposed a scenario where baryogenesis takes place in a thermal domain wall surrounding
small primordial black holes with temperature higher than the electroweak critical temperature $T_{W} \simeq 100$ GeV.
Although the idea in \cite{Nagatani} is attractive it did not received much interest. One disadvantage is the
assumption that the universe should pass from a black hole dominated era after inflation. This is not very natural,
although not forbidden, in the context of 4-dimensional cosmology \cite{khlopov2}. However, it becomes very possible in a universe
with early high energy modifications, as e.g. in a brane world
cosmology where the creation of PBHs is much more easier due to the low 5-dim Planck scale. Another
disadvantage is the final outcome that only for very large CP violating phases of order one it is possible to produce
the required baryon asymmetry and this is true for black holes masses around 100Kgs.

In the present work, first we correct a wrong constraint that was used in \cite{Nagatani} and we find that the
allowed parameter space for baryogenesis is improved. Second, the present paper studies this
baryogenesis mechanism in the context of Randall-Sundrum (RS) brane cosmology \cite{RS} and explains how it is
possible to get very easily efficient generation of baryon asymmetry even for very small CP violating angles. The allowed
by the mechanism BH mass range includes the black hole masses around the higher dimensional Planck mass. The latter is
important since this mass spectrum is energetically favorable to be generated from high energy interactions in the very
early braneworld cosmic history. Furthermore, the black hole domination era can now be naturally realized due to the
accretion in the high energy regime.

Let us explain in more detail the proposed scenario. The existence of extra dimensions \cite{DAD} is considered possible
and a lot of research has been carried out towards higher dimensional cosmological models.
Not only the cosmic geometry, but also the properties of black holes in theories with large or infinite extra dimensions
are different, since now the fundamental Planck mass is much lower.
 The proposed scenario of electroweak baryogenesis concerns the baryon asymmetry generation at the domain wall
 around annihilating PBHs in a universe with extra dimensions and in particular
 RS-II cosmology \cite{RS}. There are various mechanisms for generating these PBHs. After their formation a part of the
 universe consists of PBHs and the rest consists of radiation at temperatures lower than the electroweak scale. Soon
 after their formation PBHs start to accrete and evaporate. Depending on the accretion efficiency the two phenomena
 can dominate each other. The black hole Hawking radiation emitted thermalises the surrounding region
 at temperatures above the electroweak scale. This results to the creation of a domain wall that connects the two
 different vacua. As the Hawking particles pass this domain wall experience a CP violation leaving a net baryon
 asymmetry in the outgoing emitted radiation. At the end of PBHs' complete evaporation the Universe has been reheated
 from this Hawking flux at temperatures above the nucleosynthesis scale. The produced baryogenesis is greatly enhanced
 due to the extended lifetime of the PBHs. The prolonged lifetime is caused by the accretion factor which holds at the
 high energy cosmic period. In addition, the significant black hole accretion that takes place allows a black hole
 dominated cosmic era which helps the mechanism.

 Electroweak baryogenesis takes place at the domain wall via the standard sphaleron process \cite{CKN}. Note that the
 existence of the symmetric region surrounding the black hole washes out any baryon number created in a
 prior epoch. The proposed scenario satisfies the Sakharov's three criteria for baryogenesis \cite{Sakharov},
 \cite{Nagatani}. First, the sphaleron process that takes place at the domain wall is a baryon number violating mechanism.
 Second, although the Standard Model is a chiral theory which incorporates C-asymmetry, this is not large enough. Thus,
 we assume a two-Higgs doublets extension of the Standard Model \cite{zarikas1}, \cite{zarikas2} as the
 background field theory because it provides large CP violating phases on the Higgs sector. Finally, the outgoing
 radiation of the black hole is a non-equilibrium process.
A main advantage of this scenario, compared to the CKN electroweak baryogenesis, is the type of the phase
transition needed. At the CKN model a first order transition is required. In the present work the domain wall is created
by the thermal radiation of the black hole, and so, the phase transition can be of second order \cite{Nagatani}.
The most important result of this study is the achievement of  the observed value of
$b/s \simeq 6\times 10^{-10}$ for very small CP violating angles.

\section{Baryogenesis in the standard 4-dim FRW universe}

As mentioned in Introduction, the possibility of electroweak baryogenesis by
small primordial black holes in the standard 4-dimensional FRW universe was shown in \cite{Nagatani}. In order to
calculate the baryon-to-entropy ratio $b/s$, the author used for
the density of the black holes the Einstein equation for flat universe with matter dominant
\begin{equation}
\rho_{BH} \; = \; \frac{1}{6 \pi} \frac{m_{Pl}^2}{t^2}\,.
\end{equation}
Although, the black hole dominated era can be described by $\rho_{BH} \; \propto \; a ^{-3} $, it is not correct to fix
the unknown integration constant, and so this equation is correct up to an unknown prefactor. Indeed, we can not
normalise to the present cosmic density since black holes completely evaporate. In addition, it is not possible to
determine the prefactor using an initial black hole density at some initial time since both these quantities are unknown
and model dependent. In the discussed scenario the initial cosmic density $\rho_{BH}$ at formation is a free
parameter since it depends on the details of the black holes generation (inflation
or other mechanism). The correction of the mistake means that there is one less constraint for the black hole mass and
the scenario becomes more attractive.

Let us now correctly estimate the amount of the produced baryon-to-entropy ratio.
The total baryon number  created in the lifetime of a black hole is
\begin{equation}
B \; = \; \frac{15}{4 \pi^3 g_*} \mathcal{N} \kappa\, \alpha^{5}_{W}\, \epsilon\,\Delta\varphi_{CP}\,
\frac{m_{pl}^2}{T_{BH}T_{W}}\, ,
\end{equation}
where $g_{\ast}\simeq 100$ is the number of degrees of freedom that a BH can decay into at the electroweak
temperature, ${\cal N} \simeq O\left(1\right)$ is a model dependent
constant which is determined by the type of spontaneous electroweak baryogenesis scenario and the fermion content,
$\kappa \simeq O\left(30\right)$ is a numerical constant expressing the strength of the sphaleron process \cite{Sph},
$\alpha_W=1/30$ \cite{ASY}, $\epsilon\simeq 1/100$, $T_W=100$ GeV is the electroweak scale, and $\Delta\varphi_{CP}$
is the CP violating angle.
The total baryon number density created from all black holes is given by $ b = B n_{BH} $, where
$ n_{BH} = \frac{\rho_{BH}}{m_{BH}} $
is the number density of the black holes assuming a monochromatic spectrum of black holes. The universe after the creation
of PBHs is black hole dominated with density $\rho_{BH}$. Soon after black holes almost instantaneous evaporation, the
universe is reheated, its density has the form of radiation and is equal to $\rho_{rad} \left( t_{reh} \right)$. Thus,
\begin{equation}
\rho_{BH}(t_{reh}^{-})  \simeq \rho_{rad} \left(t_{reh} \right) = \frac{\pi^2}{30}g_{reh}T_{reh}^4 \,.
\end{equation}
The fact that $\rho_{BH}$ is a free parameter allows a freedom on the choice of $T_{reh}$. However, $T_{reh}$ has to
be below $T_{W}$ in order the baryogenesis scenario under discussion to be viable. Note that even if the scenario was
working giving finally a baryon asymmetry at reheating temperature larger than $T_{W}$, this asymmetry would be washed
out later when the universe will experience the electroweak transition. In addition, $T_{reh}$ has to be also larger
than the nucleosynthesis temperature.
Finally, estimating the cosmic entropy density as $s = \frac{2\pi^2}{45}g_{reh}T_{reh}^3 $ \cite{Kolb}, we can calculate
the total baryon-to-entropy ratio asymmetry.

Choosing $ T_{reh} = 90$ GeV  it is possible to calculate the total baryon-to-entropy ratio from all black holes
\begin{equation} \label{amount}
\frac{b}{s} \; = \; 2.4 \times 10^{-10} \;\Delta\varphi_{CP} \, .
\end{equation}
It is obvious that the required for nucleosynthesis amount of
baryon asymmetry is achieved only for $\Delta\varphi_{CP}=\pi$. Therefore, the mechanism can hardly provide sufficient
baryon asymmetry. For smaller values of  $T_{reh} $ the baryon asymmetry is further reduced. Eq. (\ref{amount})
surprisingly does not depend on the black hole mass. Although the black hole mass does not determine the baryon
asymmetry, it is constrained from two requirements regarding the existence of thermal stationary domain wall and
the black hole lifetime in comparison with the domain wall time scale. These two constraints remain the same as in
\cite{Nagatani} and give the following range for the initial BH mass
\begin{equation}
4.3\times 10^{28} \,\textrm{GeV}<m_{BH}<1.1 \times 10^{32}\, \textrm{GeV} \,.
\end{equation}

\section{Baryogenesis in the braneworld and black hole mass constraints }\label{DW}
In the framework of a RS-II braneworld embedded in a AdS bulk, primordial black holes are produced after the end
of inflation or at the very beginning of a non-inflationary flat model. Our scenario does not depend on the details
regarding the origin of the PBHs, thus this work does not study this issue. It is well known that the properties of the
brane black holes are modified compared to those in a standard cosmology \cite{Guedens}, \cite{Majumdar}. They are colder
and live longer. More important, the accretion of material from the neighborhood of the black hole can be stronger than
evaporation during a high-energy regime. This can not occur in the four-dimensional case. The presence of strong
accretion has two advantages. First, it can lead to black hole dominated universe, and second, it leads to an extension
of the black holes lifetime.

Brane black holes involved in the proposed mechanism are small enough to ensure that Hawking temperature $T_{BH}$ is
much greater than the electroweak critical temperature $T_{W}$, and so, all kinds of Standard Model (SM) particles emitted
on the brane are in the symmetric phase.  There is also emission towards the bulk, although much less, where only
gravitons are assumed to radiate. Being interested in baryogenesis, we have to deal with the emission on the brane. The
emission on the brane causes the thermalization of the surrounding region which contains radiation at low temperature.
Local thermalization applies to a region where particles have a mean free path (MFP) smaller than the size of this region.
Thus, a local temperature $T(r)$ can be defined and the mean free path of a particle $f$ is given by
$\lambda_f(T) = \frac{\beta_f}{T}$, where $\beta_f$ is a constant depending on the particle species only. The quarks
and the gluons have a strong interaction and they have the shortest MFP with $\beta_s \simeq 10$.

For a black hole with temperature $T_\BH$ there is always a closely neighborhood surrounding the horizon, which is not
thermalized, with depth $\lambda_{s}$. Moreover, in our case, the 5-dim Schwarzschild radius is much smaller than
$\lambda_{s}$, i.e.
$r_\BH  = \frac{1}{2\pi} \frac{1}{T_\BH} \ll \lambda_{s}  $.
This expression for the black hole radius $r_\BH$ in terms of the temperature $T_\BH$ is the 5-dim one. The radius and
the area of the black hole that will be used in the estimations are given by \cite{empa}
\begin{eqnarray}
&&r_{BH} \; = \; \sqrt{\frac{8}{3\pi}} \,\frac{m_{BH}^{1/2}}{m_5^{3/2}}\\
&&A_{BH} \; = \; 2\pi^2 r_{BH}^3\,,
\end{eqnarray}
which are valid provided $r_\BH<<l$. The quantity $m_5$ is the 5-dim fundamental Planck mass and $l$ is the AdS radius.
Due to the non-thermalization of the close neighborhood, the radiative particles propagate freely therein. This is why
most particles radiated do not return to the black hole, and so, the flux of the Hawking radiation obeys the
Stefan-Boltzmann's law without corrections. The outer region that is thermalized \cite{Nagatani} has as boundary the
sphere with radius $r_{o} = r_{BH}+\lambda_{s}\simeq \lambda_{s}$, which is a function of the local temperature
$T_{o}=\frac{\beta_s}{r_o}$.
The radius $r_{o}$ is the minimum thermalized radius and the temperature $T_{o}$ is the boundary temperature.

Now we consider the transfer equation of the energy in the thermalized region to determine the temperature distribution
$T(r)$ assuming the diffusion approximation of photon transfer at the deep light-depth region \cite{Mihalas}. The energy
diffusion current in Local Temperature Equilibrium (LTE) is $J_\mu
= - \frac{\beta}{3\;T(r)} \: \partial_\mu \rho$. In our case, the radiation density is
$\rho=\frac{\pi^2}{30} g_{*\SM} T^4(r)$, where $g_{*\SM} \equiv \sum_f g_{*f} = 106.75$ is the
massless freedom for all particles on the brane which is approximately equal to the massless freedom in SM. Also
$\beta/T$ is the effective MFP of all particles
by all interactions on the brane with $\beta \simeq 100$. The transfer equation is
$\frac{\partial}{\partial t} \rho = - \nabla_\mu J^\mu$. It is possible to find a stationary spherical-symmetric
solution \cite{Mihalas} which is
\begin{equation}
T(r) = \left[ T_{br}^3 + (T_{o}^3 - T_{br}^3) \frac{r_{o}}{r} \right]^{1/3}\, .
\end{equation}
This solution assumes that the freedom of the massless particles
$g_{* SM}$ is approximately constant which is valid up to all the interesting region of the domain wall.
$T_{br} = T(r \rightarrow \infty)$ is the background brane temperature. This temperature $T_{br}$ is brane model
dependent and it is related to the specific mechanism that created primordial
black holes that dominated the universe. The value of $T_{br}$ can be as large as a temperature somewhat lower that
$T_{W}$, where sphaleron rate is suppressed, and as low as zero. A very small $T_{br}$ can be realized in a particular
PBHs production model or in a scenario where the continuous accretion of PBHs made the background almost empty.

The outgoing diffusion flux is
\begin{equation}
{\cal F} =4\pi r^2 J(r)\simeq
\frac{8\pi^3}{135}\beta_s\beta \,{g_{*\SM}}\;[1 - (T_{br}/T_{o})^3]\;T_{o}^2\, .
\end{equation}
This flux must be equal to the flux of the Hawking radiation
\begin{eqnarray}
\cal{F_\BH}
&=&
4\pi r_\BH^2 \times \frac{\pi^2}{120} {g_{*\SM}}T_\BH^4 + 2\pi^2 r_{BH}^3 \times \zeta g_{bulk} T_\BH^5\, ,
\end{eqnarray}
where $\zeta$ is a constant. The radiation towards the bulk can be neglected \cite{Guedens, Kanti} because the
five-dimensional flux is negligible due to the small value of $g_{bulk}$.
The relation ${\cal F}_\BH = {\cal F}$ gives the temperature $T_{o}$ of the minimum thermalized sphere with
radius $r_{o}$
\begin{eqnarray}
r_{o} &=& \frac {16 \pi} {3} ( \beta_{s}^3\beta ) ^{1/2} [1 - (T_{br}/T_{o})^3]^{1/2}\frac{1}{T_\BH}
\end{eqnarray}
and
\begin{eqnarray}
T_o &=& \frac{3}{16\pi\sqrt{\beta_s \beta }} \; [1 - (T_{br}/T_o)^3]^{-1/2}\;T_\BH \, .
\end{eqnarray}
Assuming that $T_{br} \ll T_o$ the spherical thermal distribution
surrounding the black hole is
\begin{eqnarray}
T(r) &=& \left(
      T_{br}^3 + \frac{9}{256\pi^2}\frac{1}{\beta}\frac{T_\BH^2}{r}
     \right)^{1/3}
\label{ghw}
\end{eqnarray}
for $r > r_{o}$.


Now we will discuss the formation of a domain wall around the black hole. A two-Higss doublet model will be assumed
since it is a quite general model that can include the supersymmetric Higgs sector. The symmetry is restored at the close
neighborhood of the black hole because of its high temperature. At a greater distance, the temperature falls below
the electroweak scale and the vacuum is broken.
Therefore, an electroweak domain wall forms around the black hole, starting at radius $r_{DW}$. The presented mechanism
does not depend on the phase transition order and so it does not have to be first order. A second order transition,
which is more favorite, has been adopted. The vacuum expectation value (vev) of Higgs doublets depends on the distance
$r$ from the center of the black hole.


The two-Higgs scalar potential can be written as follows \cite{sher}
\begin{eqnarray}
V_{{\rm Higgs}} &=&\mu _{1}^{2}\Phi _{1}^{\dagger }\Phi _{1}+\mu
_{2}^{2}\Phi _{2}^{\dagger }\Phi _{2}+\lambda _{1}(\Phi _{1}^{\dagger }\Phi
_{1})^{2}+\lambda _{2}(\Phi _{2}^{\dagger }\Phi _{2})^{2}+\ \lambda
_{3}(\Phi _{1}^{\dagger }\Phi _{1})(\Phi _{2}^{\dagger }\Phi _{2})  \nonumber
\\
&&+\lambda _{4}(\Phi _{1}^{\dagger }\Phi _{2})(\Phi _{2}^{\dagger }\Phi
_{1})+{\frac{1}{2}}\lambda _{5}[(\Phi _{1}^{\dagger }\Phi _{2})^{2}+(\Phi
_{2}^{\dagger }\Phi _{1})^{2}]+V_{{\rm D}}\,,  \label{pote}
\end{eqnarray}
where $\lambda _{i}$ are real numbers and $\Phi_{1}^{\top}=
\left(\phi_{1}+i\phi_{2},\ \phi_{3}+i\phi _{4}\right)$,
$\Phi_{2}^{\top}=\left( \phi_{5}+i\phi_{6},\ \phi_{7}+i\phi_{8}\right)$.
Here, both the weak isospin doublets have weak
hypercharge $Y_{{\rm weak}}=+1$. We follow the notation of
\cite{sher} in which both Higgs doublet fields have same hypercharge.
The above potential, with the exception of $V_{\mathrm{D}}$ which we discuss
in the following, is the most general one satisfying the following discrete symmetries
\begin{equation}
\Phi _{2}\rightarrow -\Phi _{2},\;\;\Phi _{1}\rightarrow \Phi
_{1},\;\;d_{R}^{i}\rightarrow -d_{R}^{i},\;\;u_{R}^{i}
\rightarrow u_{R}^{i}\,,  \label{disc}
\end{equation}
where $u_{R}^{i}$ and $d_{R}^{i}$ represent the right-handed weak
eigenstates with charges ${\frac{2}{3}}$ and $-{\frac{1}{3}}$ respectively.
All other fields involved remain intact under the above discrete symmetries.
These symmetries force all the quarks of a given charge to interact with only
one doublet. Thus, Higgs mediated flavour changing neutral currents are
absent. If the discrete symmetry is broken during a cosmological phase
transition, it produces stable domain walls via the Kibble mechanism \cite
{kibble}. This problem can be solved by adding terms which break this
symmetry, providing at the same time the required explicit CP violation
for baryogenesis. The most general form of that part of the potential which
breaks this discrete symmetry is
\begin{equation}
V_{{\rm D}}=-\mu _{3}^{2}\Phi _{1}^{\dagger }\Phi _{2}+\lambda _{6}(\Phi
_{1}^{\dagger }\Phi _{1})(\Phi _{1}^{\dagger }\Phi _{2})+\lambda _{7}(\Phi
_{2}^{\dagger }\Phi _{2})(\Phi _{1}^{\dagger }\Phi _{2})\,+\,h.c.
\label{soft}
\end{equation}
This is commonly named as $D-$breaking part. The parameters $\mu _{3}$,
$\lambda _{6}$ and $\lambda _{7}$ are in general complex numbers
\begin{equation}
\mu _{3}^{2}=m_{3}^{2}\ e^{i\theta _{3}},\qquad \lambda _{6}=l_{6}\
e^{i\theta _{6}},\qquad \lambda _{7}=l_{7}\ e^{i\theta _{7}} \, , \label{eq5}
\end{equation}
providing explicit CP violation at the tree level.

In order to study the structure of the vacua we can perform an $SU(2)$
rotation that sets the vev's of the fields $\phi _{1,2,4}$ equal to zero.
Solving the system $\partial V_{Higgs}/\partial \phi _{i}=0$
implies several different stationary points. One of them is the usual
asymmetric minimum that respects the $U(1)$ of electromagnetism
\begin{equation}
\Phi _{1}={\frac{1}{\sqrt{2}}}\left( {\
{{0 \atop u}}}\right) ,
\;\;\Phi _{2}={\frac{1}{\sqrt{2}}}
\left( {\ {{0 \atop ve^{i\varphi_{t} }}}}\right) .
\label{I}
\end{equation}
In Eq.~(\ref{I}) $u,v,\varphi_{t} $ are real numbers. The phase $\varphi_{t} $ is
the explicit {\em CP} violating angle at tree level that appears due to the existence of the $D-$breaking terms.
The acceptable parameters of the model are those ensuring that the above
stationary point becomes the absolute minimum at zero temperature.

In the cosmological context it is necessary to use the finite temperature effective potential. It is now known that
most naturally the effective potential
contains small cubic in temperature contribution, and thus, the phase transition is a second order one. After shifting
the scalar fields about their expectation values the asymmetric minimum for the second doublet is
\begin{equation}
\phi_{c,2}= \langle\phi_2(r)\rangle = v \;f(r)\; e^{i\varphi(T,r)}, \label{VEV.eq}
\end{equation}
where
\begin{eqnarray}
 f(r) &=&
  \left\{
   \begin{array}{lcl}
    0 & & (r \leq r_\DW) \\
    \sqrt{1 - \left(\frac{T(r)}{T_\weak}\right)^2} & & (r > r_\DW)
   \end{array}
  \right. \label{FormFunc.eq}
\end{eqnarray}
is a form-function of the wall and has a value from zero to one.

In order to define a width for our domain wall $d_\DW$ in this configuration of the Higgs vev, we have to define the
value of $f(r)$ at the end of the wall. Thus, we will introduce a parameter $\xi$ that relates $d_{DW}$ with the radius
of the symmetric region $r_\DW$.
Setting $T(r_\DW) = T_{\weak}$ in Eq. (\ref{ghw}), we find
\begin{equation}
 d_\DW =\xi \, r_\DW
  = \xi \frac{9}{256 \pi^2} \frac{1}{\beta_{br}} [1 - (T_{br}/T_\weak)^3]^{-1}\frac{T_\BH^2}{T_\weak^3} \, .
\end{equation}
We are going to distinguish two cases: $\xi=1$ which correspond to an end value $f=0.6$, and $\xi=10$ which correspond
to an end value $f=0.9$.

The structure of the electroweak domain wall is determined only by
the thermal structure of the black hole and not by the dynamics of the phase transition as in the ordinary
electroweak baryogenesis scenario (the CKN model).
In the following subsections we are going to discuss two conditions ensuring that the LTE is valid. The first
constraint is the size of the domain wall to be greater than the MFP, $ 1 < d_\DW /\lambda_{s}(T_\weak)$. The second
is the black-hole lifetime to be large enough to keep the stationary electroweak domain wall,
$1 < \tau_{BH}/\tau_{DW}$. Both these constraints refer to the case without accretion. From now on we set $T_{br}$
practically zero, which is the most expected case.

\subsection{First Constraint from Thermalization Condition without accretion}
The stationary local thermal equilibrium assumption for the scalar wall is valid when the size of the wall is
greater than the MFP.
Thus, the first constraint is $ 1 <  d_\DW/\lambda_{s}(T_\weak),$ which gives
\begin{equation}
  \left(
   \frac{3}{16\pi}   \right)^2
   \frac{\xi}{\beta_s\beta\gamma  }
   \frac{T_\BH ^2}{T_\weak ^2}
> 1\, ,
\end{equation}
where $\gamma=1 - (T_{br}/T_\weak)^3$. In the present study it is more important to find constraints on the black
hole mass. Using further the modified formula for the mass of the black hole in five dimensions
\begin{equation}
m_{BH} = \frac{3}{32\pi} \frac{m_5^3}{T_{BH}^2} \, ,
\end{equation}
the equivalent constraint is
 \begin{equation}\label{bound1}
 m_{BH} < \frac{\xi}{2} \left(\frac{3}{16\pi}\right)^3 m_5^3 \;T_W^{-2} \left( \beta_s  \beta
 \gamma \right)^{-1}  \,.
 \end{equation}
It is now easy to construct Table 1 with the allowed values of black hole masses for various values of the
fundamental Planck scale.

\begin{center}
\begin{table*} \label{tab1}
\begin{tabular}{|c|c|c|}
\hline\hline
$m_{5}$ & $\xi=1$ & \,\,\,\,$\xi=10$ \,\,\,\,
\\
\hline \hline $50$\, TeV & $m_{BH}<1.3$\, TeV & $m_{BH}<13$\, TeV
\\
\hline $100$\, TeV & $m_{BH}<10.6$\, TeV & $m_{BH}<106$\, TeV
\\
\hline $1000$\, TeV & $m_{BH}<1.06\times 10^{4}$\, TeV & $m_{BH}<1.06\times 10^{5}$\, TeV
\\
\hline $5000$\, TeV & $m_{BH}<1.33\times 10^{6}$\, TeV & $m_{BH}<1.33\times 10^{7}$\, TeV
\\
\hline $ 10000$\, TeV & $m_{BH}<1.06\times 10^{7}$\, TeV &  $m_{BH}<1.06\times 10^{8}$\, TeV
\\ \hline
\end{tabular}

\vspace{0.4cm}
{\bf{Table 1:}} Summary of the first constraint on the black hole mass for various values of $m_{5}$ and for the two
domain wall thicknesses.

\end{table*}
\end{center}

\subsection{Second Constraint from BH Lifetime without accretion}
The mean velocity of the outgoing diffusing particles at radius $r_{DW}$  is
\begin{equation}
  v_\DW=\frac{J(r_{DW})}{\rho (r_{DW})}=\Big(\frac{32\pi}{9}\Big)^2\,\beta^2\,\big[1 - (T_{br}/T_{W})^3\big]^{2}
  \,\Big(\frac{T_{W}}{T_{BH}}\Big)^2\,.
\end{equation}
The characteristic time scale for the construction of the stable electroweak domain wall \cite{Nagatani} is
\begin{equation}
 \tau_\DW \simeq \frac {r_\DW} {v_\DW}
 =
 \frac{729}{262144 \pi^4}
 \frac{1}{\beta_{br}^3 \gamma^3}
 \frac{T_\BH^4}{T_\weak^5}\,.
\end{equation}

The black hole lifetime is needed to be estimated. Assuming that in the black hole dominated universe black holes are
spatially separated enough to neglect accretion
among them, we get \cite{Guedens}
\begin{equation}
\frac{dm_{BH}}{dt} \simeq -g_{*SM} \tilde{\sigma}_4  A_{eff,4} T^4 - g_{bulk} \tilde{\sigma}_5  A_{eff,5} T^5\,,
\end{equation}
where $\tilde{\sigma}_4$ and $\tilde{\sigma}_5$ are the  4-dim and 5-dim Boltzmann constants per degree of freedom
respectively, $A_{eff,4}=4\pi r_{eff,5}^2$\,, $A_{eff,5}=2\pi^2 r_{eff,5}^3$\,, and $r_{eff,5}=2r_{BH}$ is the
effective black hole radius for black body emission.
Neglecting now the evaporation to bulk \cite{Kanti}, the black hole lifetime is
\begin{eqnarray}\label{lifetime}
\tau_{BH}=t_{evap} \simeq \tilde g^{-1} \frac{l}{l_4} \left( \frac{m_{BH}}{m_4} \right)^2 t_4\,,
\end{eqnarray}
where $m_4$ the 4-dim Planck mass, $l_{4}$ the 4-dim Planck length, $t_{4}$ the 4-dim Planck time and
\begin{eqnarray}
\tilde {g} \simeq \frac{1}{160} g_{*SM} + \frac{9 \;\zeta(5) }{32 \pi^4} g_{bulk}\,.
\end{eqnarray}
We have assumed that the degrees of freedom on the thermalised region of the brane are practically the same with the
SM $g_{*SM}=106.75$ because of the high temperature of the black hole, while $g_{bulk}$ is very small and can be
ignored.

In order the mechanism to be viable the black hole lifetime should be larger than the time for the domain
wall construction. Thus, the second constraint $1 < \tau_{BH}/\tau_{DW}$ has to be respected, which gives
\begin{eqnarray}\label{bound2}
m_{BH} > \tilde g^{1/4} \frac{6561^{1/4}}{2^7 \pi^{3/2}} \left( \beta \gamma \right)^{-3/4} m_5^{9/4} T_W^{-5/4}\,.
\end{eqnarray}
This black hole mass refers to the initial black hole mass created, while $t_{evap} $ in Eq. (\ref{lifetime}) is the
time for the complete evaporation of this initial black hole mass. The bound of the first constraint also refers to
the initial black hole mass.
The above constraint is the most strict one. It can be naturally relaxed if the black hole is allowed to accrete plasma
from its neighborhood. This will be studied in next section.

For various values of the 5-dim Planck mass the black hole mass is bounded from below, as shown in Table 2.
\begin{center}
\begin{table*}[h!] \label{tab2}
\begin{tabular}{|c|c|}
\hline\hline
$m_{5}$ & Black hole mass bound
\\
\hline \hline $50$\, TeV & $m_{BH}>42.9$\, TeV
\\
\hline $100$\, TeV & $m_{BH}>204$\, TeV
\\
\hline $1000$\, TeV & $m_{BH}>3.6\times 10^{4}$\, TeV
\\
\hline $5000$\, TeV & $m_{BH}>1.35\times 10^{6}$\, TeV
\\
\hline $ 10000$\, TeV & $m_{BH}>6.45\times 10^{6}$\, TeV
\\ \hline
\end{tabular}

\vspace{0.4cm}
{\bf{Table 2:}} Summary of the second constraint on the black hole mass for various values of $m_{5}$.

\end{table*}
\end{center}

\subsection{Efficient Baryogenesis without accretion}\label{Bar}

Now we are going to estimate first the baryonic number created by a single black hole and then the cosmic baryon
to entropy ratio $b/s$. The demand $b/s \simeq 10^{-10}$ gives a strict test for all baryogenesis mechanisms.

The sphaleron process works in all the symmetric region and the domain wall. However, the baryon asymmetry production
happens in the domain wall where both CP violation and non-equilibrium conditions exist.
Futhermore, we want $f(r)=|\langle\phi_{2}(r)\rangle|/v \leq \epsilon =1/100$ in order the exponential factor in the
sphaleron
process to be of order one (otherwise the baryon asymmetry would be suppressed). This means that the effective region
of baryon generation is the region of the domain wall with small values of the Higgs scalar. The working region that
produces baryons are from $r_{DW}$ till $r_{DW}+d_{sph}$. Then $d_{sph}$ is defined from
$f(r_{DW}+d_{sph})=\epsilon$. One now can see that
$ \int_{r_\DW}^{r_\DW+d_\sph} dr \: \frac{d}{dr} \varphi(r)=\epsilon\,\Delta\varphi_\CP $, where
$\varphi(r,T)=[f(r)-1]\Delta\varphi_\CP $ \cite{CKN}. Thus,
\begin{eqnarray}
 \dot{B}
 &=&  V \; \frac{\Gamma_\sph}{T_\weak} \; {\cal N} \dot{\varphi}
       \nonumber\\
 &=&  4\pi {\cal N} \kappa\: \alpha_\weak^5 T_\weak^3 \;
       r_\DW^2 \; v_\DW
       \int_{r_\DW}^{r_\DW+d_\sph} dr \: \frac{d}{dr} \varphi(r)
       \nonumber\\
 &=&  \frac{1}{16\pi} \: {\cal N} \kappa\: \alpha_\weak^5 \:
       \epsilon\:\Delta\varphi_\CP \:
       \frac{T_\BH^2}{T_\weak}
       \nonumber\\
 &=&  A T_\BH^2, \label{BRate}
\end{eqnarray}
where $\Gamma_{sph}$ is the sphaleron transition rate, $\Delta\varphi_\CP$ the net CP phase, and it was set
for convenience $ A= \frac{1}{16\pi} \: {\cal N} \kappa\,\alpha_\weak^5 \:\epsilon\Delta\varphi_\CP
\:\frac{1}{T_\weak} $\,.
The black hole temperature can be expresses as a function of its lifetime \cite{Guedens}
\begin{eqnarray}
\TBH = \sqrt{\frac{3}{32\pi}} \,\tilde g^{-1/4} m_5^{3/4} t_{evap}^{-1/4}\,.
\end{eqnarray}
If we substitute $T_{BH}$ in Eq. (\ref{BRate}) we get
\begin{eqnarray}
\dot{B}
&=& A \frac{3}{32\pi} \tilde g^{-1/2} m_5^{3/2} \tau_{BH}^{-1/2}
    \nonumber\\
&=& \tilde A (t_{evap,0} - t)^{-1/2}\,,
\end{eqnarray}
where $ \tilde A = A \frac{3}{32\pi} \tilde g^{-1/2} m_5^{3/2} $ and  $t_{evap,0}$ is the time length for complete
evaporation of the initial black hole mass. The time $t$ runs from 0 (black hole creation) to $t_{evap,0}$, and the
baryon number created by a black hole in its lifetime is
\begin{eqnarray}
 B &=& \int_0^{t_{evap,0}} \: \dot{B} \: dt \nonumber\\
   &=& 2 \tilde A \,t_{evap,0}^{1/2}\,\,\,.
\end{eqnarray}
Using Eq. (\ref{lifetime}) to substitute $t_{evap,0}$ in terms of the initial black hole mass $\mBH$, and
$l/l_4=(m_4/m_5)^3$ we find
\begin{eqnarray}\label{B}
B &=& 2 \tilde A \:\tilde g^{-1/2}\: t_4^{1/2} \:\frac{m_{BH}}{m_4} \:\Big(\frac{l}{l_4}\Big)^{1/2} \nonumber \\
  &=& 2 \tilde A \:\tilde g^{-1/2}\: m_5^{-3/2} \:\mBH \nonumber \\
  &=& \frac{3}{(16\pi)^2}\: {\cal N}\: \kappa\: \alpha_W^5 \:\tilde g^{-1}\: T_W^{-1} \epsilon\:
  \Delta \varphi_{CP}\: \mBH\, .
\end{eqnarray}
Finally, the total baryon number density created from all black holes is $ b = B n_{BH} $,
where $ n_{BH} = \frac{\rho_{BH}}{\mBH} $ is the number density of the black holes assuming a monochromatic spectrum
of black holes.

In our scenario the universe is black hole dominated. Soon after the black holes complete evaporation, the universe
is reheated. Therefore, the cosmic black hole density just before the final stage of very rapid evaporation is almost
equal to the cosmic density radiation of the reheated plasma after the end of the evaporation
\begin{equation}
\rho_{BH}(t_{reh}^{-}) \simeq \rho_{rad} \left(t_{reh} \right) = \frac{\pi^2}{30}g_{reh}\,T_{reh}^4 \,.
\end{equation}
The entropy density is given by $s = \frac{2\pi^2}{45}g_{reh}T_{reh}^3$ \cite{Kolb}, where $g_{reh}$ is
the massless degrees of freedom of the reheated plasma in the asymmetric phase.  We choose
$ T_{reh} = 95$ GeV in order to avoid the produced baryon asymmetry to be washed out ($T_{reh}<T_W \simeq 100$ GeV).
In our study there is the freedom to select the reheating temperature in contrast to the work \cite{Nagatani}, where
mistakenly the reheating temperature was fixed using a wrong estimate of the black hole energy density. Thus, the
baryon-to-entropy ratio is
\begin{eqnarray}\label{bovers}
\frac{b}{s} &=& \frac{9}{\left(32\pi\right)^2} {\cal N} \kappa \alpha_W^5 \tilde g^{-1} \frac{T_{reh}}{T_W}
\epsilon \Delta \varphi_{CP}\,.
\end{eqnarray}
Notice that the value of the baryon-to-entropy ratio depends neither to $m_5$ nor to $m_{BH}$. For some
indicative values of $\Delta \varphi_{CP}$ the baryon-to-entropy values are
\begin{eqnarray}
\Delta \varphi_{CP} = \pi \,\,\Rightarrow\,\, \frac{b}{s} = 1.2\times10^{-10} \nonumber\\
\Delta \varphi_{CP} = 0.1 \,\,\Rightarrow \,\,\frac{b}{s} = 4\times10^{-12} \, .
\end{eqnarray}
The produced baryon-to-entropy value gets close to the observed $b/s = 6\times10^{-10}$, but only for the maximum
and not
likely $\Delta \varphi_{CP} = \pi$.

\section{Baryogenesis and constraints with accretion}\label{Accr}

In this section we will investigate the role of accretion first to the successful and efficient baryogenesis and
second to the realization of the existence of a black hole dominated era.
We assume that black holes, after their formation, not only emit but also absorb radiation from their neighborhood.
At the high energy regime of the RS universe, accretion is intense, and so, it is expected to result to a period that
the whole density of the universe is equal or close to that of the black holes. Later on, during the cosmic evolution,
evaporation starts to be more significant and finally the black holes annihilate, reheating the universe.

A phenomenological way to handle accretion is to introduce an effective factor $\textsl{f} > 1$ which denotes how
much longer becomes the lifetime of the black hole
\begin{equation}
\tau_{BH} = \textsl{f} \,\tilde g^{-1} \:m_5^{-3} \:m_{BH}^2.
\end{equation}
Now, the produced baryon number is modified to
\begin{equation}\label{Baccr1}
B = \frac{3 \textsl{f}^{\,1/2}}{(16\pi)^2} \:{\cal N}\: \kappa \:\alpha_W^5 \:\tilde g^{-1} \:T_W^{-1}
\:\epsilon \:\Delta\varphi_{CP}\: \mBH
\end{equation}
and $b/s$ of Eq. (\ref{bovers}) is multiplied by $\textsl{f}^{\,1/2}$.
In order to have $b/s \simeq 6\times 10^{-10}$, it must be
\begin{eqnarray}
\Delta \varphi_{CP} &=& 1 \,\,\Rightarrow\,\, \textsl{f} \simeq 2 \times 10^2 \nonumber\\
\Delta \varphi_{CP} &=& 0.1 \,\,\Rightarrow \,\,\textsl{f} \simeq 2\times 10^4 \nonumber\\
\Delta \varphi_{CP} &=& 0.01 \,\,\Rightarrow \,\,\textsl{f} \simeq 2\times 10^6\,.
\end{eqnarray}
As it will be more clear below, such values of $\textsl{f}$ can naturally be realised. This is a remarkable result. It is
very easy to produce large values of baryon asymmetry and even larger than the required amount, for very small values of
$\Delta \varphi_{CP}$.

Let us discuss at this point the various constraints in the presence of accretion that extends the black hole
lifetime. The first constraint Eq. (\ref{bound1}) remains intact and refers to the maximum value of black hole mass
reached just before
the evaporation start to dominate the accretion. However, the second bound is modified. It becomes less strict because
the black hole lifetime is lengthened. The constrained black hole mass refers to the initial value of the black hole mass
\begin{eqnarray}
m_{BH,i} > \textsl{f}^{\,-1/4} \:\tilde g^{1/4} \:\frac{6561^{1/4}}{2^7 \pi^{3/2}} \left( \beta\gamma
\right)^{-3/4} \:m_5^{9/4} \:T_W^{-5/4}\, .
\end{eqnarray}
Table 3 shows some black hole mass bounds from below for some representative combinations of the involved
free parameters.

In summary, taking into consideration both first and second constraint and demanding $b/s \simeq 6\times 10^{-10}$, we
can find allowed black hole mass ranges for various values of $m_{5}$ and $\Delta \varphi_{CP}$, i.e. for
$m_{5}=50\:$TeV
and $\Delta \varphi_{CP}=0.01$ the allowed range is $1.1\:$TeV$<m_{BH}<13\:$TeV, for $m_{5}=100\:$TeV and
$\Delta \varphi_{CP}=0.01$ the allowed range is $5.3\:$TeV$<m_{BH}<106\:$TeV, etc. These ranges differ from the
previously mentioned case without accretion. There, the estimated ranges show the allowed range of the initial
black hole mass. Here, in this section that accretion is added, the estimated ranges show the allowed wider possible
range of the time dependent black hole mass during the accretion period. Thus, the range $5.3\:$ TeV$<m_{BH}<106\:$ TeV
means that the initial black hole mass can be as low as 5.3 TeV and increases during accretion as large as 106 TeV.
It is worth mentioning that for smaller values of CP, which is more favourable, the above allowed black hole mass
ranges enlarge!

\begin{center}
\begin{table*}[h!] \label{tab3}
\begin{tabular}{|c|c|c|}
\hline\hline
$m_{5}$ & $\textsl{f}$ & \,\,\,\,Initial black hole mass \,\,\,\,
\\
\hline \hline $50$\, TeV & $2\times 10^2 \left( \Delta \varphi_{CP} = 1 \right)$ & $m_{BH,i} > 11\:$TeV
\\
\hline $50$\, TeV &  $2\times10^4 \left( \Delta \varphi_{CP} = 0.1 \right)$ & $m_{BH,i} > 3.5\:$TeV
\\
\hline $50$\, TeV & $2\times10^6 \left( \Delta \varphi_{CP} = 0.01 \right)$ & $m_{BH,i} > 1.12\:$TeV
\\
\hline $100$\, TeV & $2\times10^4 \left( \Delta \varphi_{CP} = 0.1 \right)$ & $m_{BH,i} > 17\:$TeV
\\
\hline $ 100$\, TeV & $2\times10^6 \left( \Delta \varphi_{CP} = 0.01 \right)$ &  $m_{BH,i} > 5.3\:$TeV
\\ \hline
\end{tabular}

\vspace{0.4cm}
{\bf{Table 3:}} Summary of the second constraint on the initial black hole mass for various values of $m_{5}$ and
$\textsl{f}$. The values of $\textsl{f},\,\Delta \varphi_{CP}$ are those that give the observed baryon asymmetry ratio.
\end{table*}
\end{center}

In the RS model, there is a characteristic transition time $t_c$ that denotes the passage from the high-energy regime
with the unconventional Hubble law to the low-energy regime. An interesting and workable case is when accretion is
stronger than evaporation and continues till $t_c$, while afterwards evaporation is the dominant term in the differential
equations. This case has been discussed in \cite{Guedens}. The black hole lifetime is
\begin{eqnarray}
\tau_{BH} &=& t_c + \tilde g^{-1} m_5^{-3} m_{BH,max}^2 \nonumber\\
          &=& \frac{1}{2} \frac{m_4^2}{m_5^3} + \tilde g^{-1} \frac{m_{BH,max}^2}{m_5^3}\,,
\end{eqnarray}
where $m_{BH,max}$ is now the black hole mass at $t_c$.
For $m_{BH,max} < m_4$, which is always the case in the present study, it is $\tau_{BH} \simeq t_c$ and so the baryon
number produced by a black hole becomes
\begin{eqnarray}\label{Baccr2}
B &=& 2 \tilde A t_c^{1/2} = 2 A \frac{3}{32 \pi} \tilde{g}^{-1/2} m_5^{3/2} t_c^{1/2} \nonumber\\
  &=& \frac{3/\sqrt{2}}{\left( 16 \pi \right)^2} \tilde g^{-1/2} {\cal N} \kappa \alpha_w^5 \epsilon \Delta
  \varphi_{CP} T_W^{-1} m_{4}\,.
\end{eqnarray}
Note that the final baryon to entropy ratio does not depend on $m_5$. Some indicative combinations of required CP
angles and black holes masses ensuring $b/s \simeq 6\times 10^{-10}$ are
\begin{eqnarray}
m_{BH,max} &=& 10^4 TeV \,\,\Rightarrow \,\,\Delta \varphi_{CP} = 10^{-11} \nonumber\\
m_{BH,max} &=& 10 TeV \,\,\Rightarrow \,\,\Delta \varphi_{CP} = 10^{-14} \nonumber\\
m_{BH,max} &=& 1 TeV \,\,\Rightarrow \,\,\Delta \varphi_{CP} = 10^{-15}  .
\end{eqnarray}
The qualitative behaviour of the black hole mass time evolution is very sensitive on the accretion efficiency. If the
efficiency is low, the dominant accretion stops inside the high energy regime and the above estimated $m_{BH,max}$
masses decrease.

Here, the first constraint remains the same and refers to the maximum black hole mass.
On the other hand, the second constraint is practically always satisfied since now $t_{c}$ is very large, i.e. for
$m_5=100\:$TeV,
$m_{BH}=10\:$TeV, we get $t_{c}/t_{evap}=10^{29}$. The second constraint can be estimated from
\begin{eqnarray}
&&\,\,\,\,\,\,\,\,\,\,t_{evap}+t_c > \tau_{DW} \nonumber\\
&&
\Leftrightarrow \,\,\,m_{BH,i}^2 > \frac{729}{262144 \pi^4} \frac{2}{\beta^3 \gamma^3} \left( \frac{3}{32 \pi} \right)^2
\frac{m_5^9}{m_{4}^2} \frac{1}{T_W^5}\,.
\end{eqnarray}

Since the accretion efficiency, the time at black hole formation, and initial black hole mass are unknown quantities,
it is not useful to study quantitatively and fully various cases, solving the differential equation of black hole
mass time evolution.
However, it becomes apparent that successful baryogenesis can be achieved for very small values of CP angles, which
can be provided also from different matter content than the two-Higgs model. Thus, our scenario does not depend on a specific
form of the Higgs sector. It only requires a small CP angle on one scalar vev.

Note also that some of the evaporated baryon excess could be eaten from the same black hole during accretion. This phenomenon
is expected not to be significant since the Hawking radiation has the escape velocity from the gravitational field. Another possibility is that some of the evaporated baryon asymmetric radiation to be eaten by nearby black holes. This complication becomes unimportant assuming that all black holes are initially widely separated while the expansion further increases the inter black holes distances.

\subsection{Black hole domination era due to accretion}
Let us discuss now the possibility of a black hole dominant era in the RS setup. This situation becomes easily
realized due to the
strong accretion at the high-energy regime. The differential equation that describes the black hole mass time
evolution is
\begin{equation}\label{difeqn}
\frac{dm_{BH}}{dt}=  F\pi r_{eff,5}^2 \,\rho_{rad} - g_{*SM} \frac{3 \Gamma \left(4\right) \zeta
\left(4\right)}{2^6 \pi^4}\frac{m_5^3}{m_{BH}}\,,
\end{equation}
where $F$ is the accretion efficiency and $\rho_{rad}$ is the energy density of the surrounding radiation.

According to the study in \cite{Guedens}, if the accretion efficiency factor is $F>0.78$, the PBH grows (accretion
dominates evaporation) until $t_c$ is
reached, provided the initial black hole mass is $m_{BH,i} > m_5$.
If $F<0.78$ then the loss due to evaporation is larger than the gain. There is also a case ($m_{BH,i} \gg m_5$ and
low efficiency) where we have more accretion
than evaporation till ``halt" time $t_h$ in the high-energy regime.
In radiation dominated high-energy regime it can be proved that $t_h^{1-q} \simeq q\left[ 1+ \left( 1-q \right)
\frac{4 \sqrt{\nu}}{\tilde{g}} \left( \frac{m_{BH,i}}{m_5} \right)^{3/2}\right]\,t_i^{1-q}$, where $q=4F/\pi$ and $\nu$
denotes what fraction of the horizon mass the initial black hole mass comprises.

Demanding a smaller evaporation than accretion, the second term in Eq. (\ref{difeqn}) becomes suppressed. The differential
equation now takes the form
\begin{equation}\label{diffRS}
  \frac{dm_{BH}}{dt}=  \frac{2\,F}{\pi} \,\frac{m_{BH}}{t}
\end{equation}
with solution $m_{BH}=m_{BH,i}(t/t_i)^{2F/\pi}$. Let us assume the case where $\rho_{rad}$ initially, at the PBHs'
formation time $t_i$, is much higher than $\rho_{BH}$, i.e. $\rho_{rad, i}=\mu_i \,n_{BH}\, m_{BH,i}$ with $\mu_i>1$.
If at the end of accretion period at $t=t_f$ we have a black hole domination, then
$\rho_{rad, f}=\mu_f \,n_{BH}\, m_{BH,i}\,(t_{f}/t_{i})^{2F/\pi}$ with $\mu_f<1$. Energy conservation implies
\begin{equation}\label{cond}
  t_f=\Big(\frac{1+\mu_i}{1+\mu_f}\Big)^{\pi/2F} \,t_i \,.
\end{equation}
This expression implies that it is always possible to start with a radiation dominated era and end in a black hole
dominated era within the high-energy regime. This holds since choosing a small enough value of $t_i$, the time duration
$t_f$ can be smaller than $t_h$ or $t_c$, which is the upper bound for the dominant accretion period.

\subsection{PBHs constraints }

The model described in the present work should comply with the constraints coming from observational data.
These constraints refer to the fraction at formation time of the mass of the universe going into PBHs,
namely they refer to the quantity $\alpha_{i}=\frac{\rho_{i,BH}}{\rho_{i,tot}}$.
We summarize the possibly relevant observational constraints in relation to the very small PBH masses
appeared in our scenario \cite{carr2}, \cite{sendou}, \cite{tashiro}, \cite{Sendouda}, \cite{khlopov1}, \cite{khlopov2}.
PBHs with lifetime smaller than $10^{-2}s$ are free from BBN constraints because they evaporate well before
weak freeze-out and leave no trace.
Observation of the extragalactic photon background provides no limit on $\alpha_{i}$ for very small PBHs
as the ones discussed here.
For PBHs with masses below $10^{4}g$, the emitted photons from the evaporation do not violate the observed
value of photon-to-baryon ratio.
Supersymmetry or supergravity relics provide no limit on $\alpha_{i}$ for very small BH masses, so that the
observed cold dark matter density is not exceeded.
If PBH evaporations leave stable Planck-mass relics, these contribute to the dark matter, and in order not to
exceed the critical density there arises an upper bound on $\alpha_{i}$, but for masses not as small as the ones here.
In general, the analysis of all the above constraints has been performed
for the standard four-dimensional cosmology, so an appropriate analysis should consider the corresponding corrections
due to extra dimensions. To conclude, all the constraints refer to four-dimensional PBHs with masses at least
$10^{-5}g$ (created at Planck time $10^{-43}s$). Since our scenario is a higher-dimensional one with a fundamental
mass scale of TeV, the allowed PBH masses are of this order, and therefore, it is quite probable that they
are too small to be constrained by observational data.

Another general issue regarding accretion of matter into a black hole is the formation or not of shock waves subject
to various conditions \cite{shock}.  A shock is formed when the rotating flow has a high angular velocity that passes
the centrifugal barrier. However, even if this velocity is somewhat lower, shocks can also be formed if the pressure of
the flow is large. Most literature analyses semi-analytically and numerically this phenomenon in the context of
astrophysical black holes. In addition, there are theoretical works assuming newtonian or post-newtonian physics
that describe analytically the existence criteria of shock waves. Typically if the angular momentum $l$ is close to
the marginally stable value and the initial kinetic energy $e$ for accretion or thermal energy for wind is within a
few percent of the rest mass energy, the flow should pass through a shock. One major problem is that for a given set
of $e$ and $l$ for every solution that includes a shock there exists another solution which is shock free. Numerical
simulations show that if there are significant perturbations in the flow of falling material, more than a certain
degree, then there is shock formation. However, these numerical works concern choice of parameters relevant
for astrophysical black holes. For primordial black holes generated in a 4-dim FRW universe a crucial criterion is that
the perturbation amplitude $\delta$ (defined as the relative mass excess inside the overdense region measured when it
had the same scale as the cosmological horizon) is greater than a threshold value $\delta_{c}$ . For perturbation
with $\delta$ close to $\delta_{c}$ numerical calculations reveal that shocks are always formed.

In our context there are primordial black holes embedded in a surrounding cold radiation bath. The radiation
temperature during the accretion period needs to be lower than the electroweak scale and it can be very much lower.
However, this temperature depends on the specific cosmic scenario that creates the primordial black holes. It is
reminded that the universe is reheated after the evaporation of all cosmic primordial black holes. Thus, the falling
material needs not to have large kinetic energy.
In addition, the fact that the accretion happens into the RS high energy regime makes the surrounding radiation plasma
to be eaten more effectively contrary to the conventional FRW, as the expansion proceeds. The reason is that the slower
decrease of the background
density during the high-energy regime makes accretion important. Therefore the high energy regime increases the
accretion efficiency and not necessarily the speed of the rotating flow of falling material. Nevertheless, a complete
study should consider i) the profile type of initial perturbations that created PBHs, ii) the five-dimensional geometry
of the black holes, iii) the small black holes masses which make them hot enough to produce significant quantum
evaporation, and iv) the complication that the escape of possible shock waves may feed the accretion of nearby black
holes, depending on the inter black holes distances and the expansion rate.
It worths investigating in a separate work for a certain cosmic scenario of production of brane primordial black
holes, the possibility of the formation of shock waves during the accretion period. The study would almost certainly
require numerical analysis.

Finally, let us finish mentioning one important point. Although we have performed our analysis of the BH accretion
in the high energy regime of a RS cosmology, a similar analysis should also hold for any early cosmology with high energy
modifications. Thus, alternative modified gravity models \cite{Harko} or even braneworld models with high
curvature corrections \cite{Germani} should in principle equally well produce significant baryon asymmetry.

\section{Black holes mass spectrum}
In this section a discussion regarding the effects of a possible initial mass spectrum of primordial black holes is
presented.
Till now, a monochromatic mass spectrum was assumed. This was necessary in order to be able to find analytical
expressions and inequalities and check first if the proposed baryogenesis mechanism works without any conflicts, and second if
it is able to generate the required amount of baryon asymmetry.

Let us discuss how the various constraints on the black hole mass are affected from the existence of a black hole
mass spectrum. The first constraint that ensures thermalization demands the size of the wall to be larger than the
mean free path and this suggests an upper bound on the black hole mass. It is obvious that all the black holes of the
spectrum with mass greater than this upper bound are not hot enough to thermalize the surrounding domain wall and
thus they do not produce any baryon asymmetry. The exact distribution of the mass spectrum and its upper tail will
determine how large or small a correction to the baryon asymmetry will be. It worths as a future work to adapt a
specific mechanism of creation of primordial black holes and analyze numerically the proposed baryogenesis scenario.
The second constraint, which comes from demanding the black hole lifetime to be larger than the time scale of stable
domain wall construction, generates a lower bound on the black hole mass. Black holes smaller than this limit
evaporate too soon. However, as we have explained previously, this constraint in the presence of dominant accretion
in the high energy regime becomes extremely weak, since the lifetime of small black holes is considerably extended.

Nevertheless, a not very narrow mass spectrum may modify the calculations of the produced baryon asymmetry.
Indeed Eqs. (\ref{B}), (\ref{Baccr1}), (\ref{Baccr2}) still hold, but now the total baryon number density created from
all black holes is given from a more complicated expression
\begin{equation}\label{Bspectrum}
   b = \int ^{\infty} _{0} B\, N(m,t)\,dm\,,
\end{equation}
with $N$ the number density of the mass spectrum of black holes with masses between $m$ and $m+dm$.
As a general conclusion it suffices to state that the very efficient baryogenesis due to accretion remains
unaffected from the presence of mass spectrum. Based on a certain cosmological scenario of creation of PBHs one
can estimate the exact baryon asymmetry straightforwardly. More details will follow concerning the relation of
the black hole mass spectrum and the time evolution of the scale factor and the cosmic densities.

We are now going to obtain the equations that determine the evolution of the spectrum of primordial black holes,
taking full account of either evaporation into radiation or accretion eating radiation, as well as the effect of
the black holes on the evolution of the scale factor. It is assumed that the number density of the initial black
hole spectrum is described by a power-law form, following \cite{Carr75}, \cite{Barrow}. Thus, the initial number
density of the primordial black hole spectrum between $m_0$ and $m_0+dm_0$ is
\begin{equation}\label{initialnumber}
  N(m_0)dm_0=A \,m_{0}^{-n}\,\Theta(m_0-m_c)\,dm_0
\end{equation}
with $m_0=m(t=0)$ the initial black hole mass. It has been assumed that all the black holes of the mass spectrum
form simultaneously at a certain time otherwise analytic results become unnecessarily hard to be obtained. The
$\Theta$ function (with $\Theta=1$ for $x>0$ and $\Theta=0$ for $x\leq 0$) is introduced in order to model the
presence of a cut-off mass in the spectrum and protects from the appearance of
divergences at low masses limit. The cut-off mass $m_c$ is natural to be a factor of the fundamental Planck
mass, $m_c=k\,m_5$, with $k$ an arbitrary dimensionless constant. The power law should be such that the total energy density does not diverge at
large masses and this implies $n>2$. However, as Carr notes \cite{Carr75}, initial density perturbations in FRW cosmologies that produce
primordial black holes suggest values in the range $2<n<3$. In RS cosmology similar ranges for the power law apply
\cite{Sendouda}. The constant $A$ represents the amplitude of the spectrum and has appropriate units such that
$N(m_0)dm_0$ is number density.

Next step is to determine analytically the spectrum $N(m,t)dm$. Both evaporation and accretion modify the value of the
cut-off mass (evaporation reduces it). The number density at a given time will be
\begin{equation}\label{initnd}
  N(t)=\int ^{\infty} _{0} N(m,t)\,dm\,,
\end{equation}
while the energy density is given by
\begin{equation}\label{energydensity}
  \varrho_{BH}(t)=\int ^{\infty} _{0} N(m,t)\,m\,dm\,.
\end{equation}
Since the purpose is to evaluate the modifications on the evolution of cosmic densities due to PBHs back-reaction, we
are going to distinguish two cases. The first case concerns the description of the cosmic evolution after the
evaporation starts to become dominant compared to the accretion. The second case is the description of the cosmic
evolution during the era when accretion mainly determines the black hole mass evolution. Any attempt to seek analytical
cosmological solutions considering both accretion and evaporation at the same time proved to be non fruitful. However,
the most realistic scenario is this that comprises a long dominant accretion time period during the RS high energy
regime which ends and is followed by a dominant evaporation era that results to a reheated radiation dominated universe.

\subsection{Dominant evaporation era}
First we will study the most interesting case when accretion has just stopped to be significant and evaporation
dominates the evolution of the black hole mass. The significance of this analysis lies on finding the modifications on the expansion rate that have to be decreasing, allowing the emergence of the conventional radiation expansion law.  Accretion has extended the black hole lifetime and thus significant
baryogenesis has already been achieved. As soon as evaporation starts to dominate, something that is expected to be
certainly true after the high energy regime $t>t_c$, the black hole mass rapidly decreases. The purpose is to estimate
deviations on the cosmic densities and scale factor time evolutions.

The black hole mass spectrum has a time evolution first due to the expansion, which will be added later, and second and more
physically important due to the evaporation. Denoting $m_{BH}$ by $m$ as above, the rate of loss of a
single black hole is given by
\begin{equation}\label{lossrate}
  \dot{m}=-g_{tot}\,\frac{m_5^{3}}{m}\,,
\end{equation}
where
\begin{equation}\label{gtot}
  g_{tot}=\frac{1}{2} \left[\frac{0.0062}{G_{brane}}g_{*SM}+\frac{0.0031}{G_{bulk}}g_{bulk}\right]\simeq g_{*SM} \frac{3\, \Gamma \left(4\right) \zeta
\left(4\right)}{2^6 \,\pi^4}
\end{equation}
and the second expression disregards the very small bulk contribution in $g_{tot}$. The quantities
$G_{brane},\,\,G_{bulk}$ represent the grey-body factors for brane and bulk respectively. In the standard cosmology
the grey-body factor is equal to 2.6, but precise values are not well known for the braneworld, see discussion in \cite{Guedens}.
The effective degrees of freedom $g_{tot}$ is a funtion of temperature. For the time periods referring to the two
cases we study in this section, we assume it is a constant. Eq. (\ref{lossrate}) can now be integrated and
gives
\begin{equation}\label{massbh}
  m^2=m_0^2-2\,g_{tot}\,m_5^3\,t\,.
\end{equation}
Solving Eq. (\ref{massbh}) with respect to $m_0$ and differentiating, we are able to find the time evolution of the
number density between $m$ and $m+dm$ at time $t$. The time evolved spectrum is
\begin{equation}\label{nd}
   N(m,t)dm=A\, m^{-n}\,\left( 1+ \frac{2\,g_{tot}\,m^3_5\,t}{m^2} \right)^{-(n+1)/2} \,\Theta(m-m_{cr}(t))\,dm\,,
\end{equation}
where now the cut-off mass has also time evolved and is given by
\begin{equation}\label{mcr}
  m_{cr}(t)= k \, m_5 \, (1-2\,g_{tot}\, k^{-2} \, m_5 \, t)^{1/2}\,.
\end{equation}
It is obvious that after a time $t_{lim}=\frac{k^2}{2\,g_{tot}\,m_5}$ the cut-off mass reaches zero.

The energy per volume that is transferred from the black hole density to the radiation between times $t$ and $t+dt$
can be determined from Eq. (\ref{energydensity}) and is given by
\begin{equation}
  dE=\varrho_{BH}(t)-\varrho_{BH}(t+dt)=-\frac{\partial \varrho_{BH}}{\partial t}\,dt\,.
\label{dE}
\end{equation}
The energy density rate can be estimated using the identity
\begin{equation}\label{identity}
\frac{d}{dx}\int ^{f(x)} _{g(x)} h(x,y)\,dy=\int ^{f(x)} _{g(x)}\frac{\partial h(x,y)}{\partial x} \,dy + h(x,f(x))
\frac{df(x)}{dx}-h(x,g(x))\frac{dg(x)}{dx}\,,
\end{equation}
thus we find
\begin{eqnarray}
  \frac{dE}{dt}&=&-\frac{d}{dt}\,\int ^{\infty} _{0} N(m,t)\,m\,dm \\
  &=&A\,g_{tot}\,(n+1)\,m^3_5\,\int ^{\infty} _{m_{c,max}} m^{-n-1}\,
  \left( 1+ \frac{2\,g_{tot}\,m^3_5\,t}{m^2} \right)^{-(n+3)/2} \,dm \nonumber\\
  &&-  A\,g_{tot}\,k^{-n} \,m^{-n+3}_5\,(1-2\,g_{tot}\, k^{-2} \, m_5 \, t)\,\Theta\Big(\frac{k^2}{2\,g_{tot}\,m_5}-t\Big)\,,
\label{dedt}
\end{eqnarray}
where
\begin{equation}\label{max}
  m_{c,max}(t)=max[0,m_{cr}(t)]\,.
\end{equation}
The first term in Eq. (\ref{dedt}) expresses the evolution of the spectrum and is the only non-zero term at late times.
The second part of Eq. (\ref{dedt}) arises due to the time evolution of the mass cut-off. It is apparent that for times
larger than $t_{lim}$ the lightest black holes completely evaporate and the $\Theta$ function causes this term to vanish.
Now it is possible to write the full equations of motion describing the expansion. We define for convenience the
scale factor at $t=0$ to be one, $a(t=0)=1$, where $t=0$ corresponds to the time of primordial black holes formation.
In all quantities calculated so far the dilution from
expansion will have to be added, i.e. the spectral amplitude $A$ becomes $A\,a^{-3}$, the comoving density is
$\rho_{BH}=\varrho_{BH}\,a^{-3}$, and the comoving energy is $E_{com}=Ea^{-3}$.
Since the case under study concerns the dominant evaporation regime which most naturally starts after the high energy
regime of the RS cosmology, the set of equations is the following
\begin{equation}\label{conservation}
  \Big(\frac{\dot{a}}{a}\Big)^2=\frac{8\pi}{3\,m_4^2}(\rho_{rad}+\rho_{BH})
\end{equation}
and
\begin{equation}\label{expansion}
  \dot{\rho}_{rad}=-4\frac{\dot{a}}{a}\rho_{rad}+\frac{dE_{com}}{dt}\,.
\end{equation}
Note that we have assumed that black holes exert unimportant kinetic pressure. Furthermore, for late times $t>t_{lim}$
\begin{eqnarray}\label{eqrho}
  \rho_{BH}&=&\frac{1}{a^3}\int ^{\infty} _{0} N(m,t)\,m\,dm \nonumber\\
  &=&\frac{A}{a^3}\int ^{\infty} _{0} m^{-n+1}\,\left( 1+ \frac{2\,g_{tot}\,m^3_5\,t}{m^2}\right)^{-(n+1)/2}\,dm\,.
\end{eqnarray}
It is convenient to set $L=2\,g_{tot}\,m_5^3$ and $\mu=\frac{m}{\sqrt{L\,t}}$. Now it is possible to estimate the integral
\begin{eqnarray}\label{eqrho2}
  \rho_{BH}&=&\frac{A}{a^3}(L\,t)^{\frac{-n+2}{2}}\int ^{\infty} _{0} \mu^{-n+1}
  \Big(1+\frac{1}{\mu^2}\Big)^{-(n+1)/2}\,d\mu \nonumber\\
  &=& \frac{A}{a^3}(2\,g_{tot}\,m_5^3\,t)^{\frac{-n+2}{2}}\,\frac{\sqrt{\pi}}{4}\,
  \frac{\Gamma (-1+\frac{n}{2})}{\Gamma(\frac{1+n}{2})}\,,
\end{eqnarray}
which holds for $n>2$. We observe that it became possible to find the power of the time evolution of the black hole
density $\rho_{BH}\propto t^\frac{-n+2}{2}$. It depends on the spectral index $n$ which most expectedly takes values
$2<n<3$. The comoving transfer rate per volume $\frac{dE_{com}}{dt}$ for $t>t_{lim}$ is given by
\begin{eqnarray}\label{decdt}
  \frac{dE_{com}}{dt}&=&\frac{A}{a^3}(n+1)\,g_{tot}\,m_5^3\int ^{\infty} _{0} m^{-n-1}\,
  \left( 1+ \frac{2\,g_{tot}\,m^3_5\,t}{m^2} \right)^{-(n+3)/2} \,dm\nonumber\\
  &=&\frac{A}{a^3}(L\,t)^{-n/2}\,(n+1)\,g_{tot}\,m_5^3\int ^{\infty} _{0} \mu^{-n-1}
  \Big(1+\frac{1}{\mu^2}\Big)^{-(n+3)/2}\,d\mu \nonumber\\
  &=& \frac{A}{a^3}\,(n+1)\,(2\,g_{tot}\,m_5^3)^{1-\frac{n}{2}}\,\frac{\sqrt{\pi}}{8}\,
  \frac{\Gamma (\frac{n}{2})}{\Gamma(\frac{3+n}{2})}\,t^{-\frac{n}{2}}\,.
\end{eqnarray}
There are deviations in the time evolution of the radiation density compared to the conventional FRW model. Actually, Eq. (\ref{expansion})
using (\ref{decdt}) can be integrated to
\begin{equation}
\rho_{rad}=\frac{c_0}{a^{4}}+\frac{c_1}{a^{4}}\int a \,t^{-n/2}dt\,
\label{sjk}
\end{equation}
where $c_0$ is an arbitrary constant and
\begin{equation}\label{c1}
  c_1=A\,(n+1)\,(2\,g_{tot}\,m_5^3)^{1-\frac{n}{2}}\,\frac{\sqrt{\pi}}{8}\,
  \frac{\Gamma (\frac{n}{2})}{\Gamma(\frac{3+n}{2})}\,.
\end{equation}
Then, Eqs. (\ref{conservation}), (\ref{eqrho2}) give an integro-differential equation for the scale factor
\begin{equation}\label{hubbleevap}
\Big(\frac{\dot{a}}{a}\Big)^{2}=\frac{\tilde{c}_0}{a^{4}}+\frac{\tilde{c}_1}{a^{4}}
\int a\, t^{-n/2}dt+\frac{\tilde{c}_2}{a^{3}}t^{\frac{2-n}{2}}\,,
\end{equation}
where $\tilde{c}_0$ is an arbitrary constant and
\begin{equation}\label{c1tilda}
  \tilde{c}_1=\frac{8\pi}{3\,m_4^2}\,c_1\,,
\end{equation}
\begin{equation}\label{c2tilda}
  \tilde{c}_2=\frac{8\pi}{3\,m_4^2}\,A\,(2\,g_{tot}\,m_5^3)^{\frac{2-n}{2}}\,\frac{\sqrt{\pi}}{4}\,
  \frac{\Gamma (-1+\frac{n}{2})}{\Gamma(\frac{1+n}{2})}\,.
\end{equation}
Now, Eq. (\ref{hubbleevap}) can be converted, after a differentiation, into a Raychaudhuri equation
\begin{equation}
2\,a^{3}\,H(2H^{2}+\dot{H})=t^{-n/2}\Big(\tilde{c}_1+\tilde{c}_2\,\frac{2-n}{2}+\tilde{c}_2\,H\,t\Big).
\label{raycha}
\end{equation}
The derived Raychaudhuri equation cannot be solved analytically but it can be shown that
\begin{equation}
a \propto t^{1/2}
\label{haj}
\end{equation}
is a solution of Eq. (\ref{raycha}) neglecting terms of order
$t^{-n/2}$. Thus, for times much after the end of evaporation the
usual expansion is recovered. More definite results can be only
extracted from numerical calculations and simulations covering
various ranges of the involved free parameters.

\subsection{Dominant accretion era}
Here we will analyse another interesting case. It refers to the time period after primordial black hole creation.
Since the creation happens in the high energy regime of RS cosmology it is expected accretion to be much more
significant than evaporation. The purpose is to estimate the time evolutions of the cosmic densities and the scale
factor.

The black hole mass spectrum has now a time evolution due to the accretion, apart from the expansion which will be
added later. The rate of loss of a single black hole is given by
\begin{equation}\label{lossrateaccretion}
  \dot{m}= F \, \pi \, r_{\text{eff},5}^2\, \rho_{rad}=F\,\frac{32}{3}\,\frac{m}{m_5^3}\rho_{rad}\,,
\end{equation}
where the $\rho_{rad}$ represents the surrounding to the black holes radiation density. The time duration of this
case, where accretion is dominant, is much longer than the regime of dominant evaporation. Since most baryon asymmetry
is produced during this accretion period it worths describing the complicated equations of motion.
Eq. (\ref{lossrateaccretion}) can be solved and gives
\begin{equation}\label{massbhac}
  m=m_0\,\zeta \, \exp\Big(\int ^{t} _{0} \rho_{rad} \,dt \Big)\,,
\end{equation}
where $\zeta=\exp(\frac{32\,F}{3\,m_5^3})$\,.

Solving Eq. (\ref{massbhac}) with respect to $m_0$ and differentiating, we are able to find the time evolution of
the number density between $m$ and $dm$ at time $t$, with the help of Eq. (\ref{initialnumber}). The time evolved
spectrum now is
\begin{equation}\label{ndac}
   N(m,t)dm=A\,\zeta^{n-1}\,exp\Big[(n-1)\int ^{t} _{0} \rho_{rad} \,dt\Big]\, m^{-n}\,\Theta(m-m_{ca}(t))\,dm\,,
\end{equation}
where the cut off mass has been time evolved from $m_c$ to $m_{ca}$ given by
\begin{equation}\label{mca}
  m_{ca}(t)= k \, m_5 \,\zeta\, \exp\Big(\int ^{t} _{0} \rho_{rad} \,dt\Big)\,.
\end{equation}
It is obvious that contrary to the previous case the cut off mass does not equal zero at any time.

The energy per volume that is transferred from the eaten radiation to the black hole density between times $t$ and
$t+dt$ can be determined from $dE=\varrho_{BH}(t)-\varrho_{BH}(t+dt)=-\frac{\partial \varrho_{BH}}{\partial t}\,dt$
and the energy density rate can be estimated using in addition Eq. (\ref{identity}). Thus
\begin{eqnarray}\label{dedtaccr}
\frac{dE}{dt}&=&A\,\frac{n+1}{n-2}\,\zeta\,k^{-n+2}\,m_5^{-n+2}\,\exp\Big(\int^{t}_{0}\rho_{rad}\,dt\Big)\,\rho_{rad}\\
 && - A\,\zeta\,k^{-n+2}\,m_5^{-n+2}\,\exp\Big(\int^{t}_{0}\rho_{rad}\,dt\Big)\,\rho_{rad}\,\Theta(m-m_{ca})\,.
\end{eqnarray}
The first term in Eq. (\ref{dedtaccr}) expresses the evolution of the spectrum, while the second part arises due
to the time evolution of the mass cut-off. In this second case this term does not vanish as long as evaporation is
less significant than accretion.
Now it is possible to write the full equations of motion describing the expansion. All densities should become
comoving multiplying them with $a^{-3}$. In this case of dominant
accretion regime we are clearly in the high energy regime of the RS cosmology. Therefore the set of equations is the
following
\begin{equation}\label{conservationaccr}
\Big(\frac{\dot{a}}{a}\Big)^2=\frac{8\pi}{3\,m_4^2}\Big(\rho_{rad}+\rho_{BH}+\frac{1}{2\lambda}
(\rho_{rad}+\rho_{BH})^2\Big)
\end{equation}
with $\lambda=\frac{3\,m_5^6}{4\pi\,m_4^2}$ and
\begin{equation}\label{expansionaccr}
  \dot{\rho}_{rad}=-4\frac{\dot{a}}{a}\rho_{rad}+\frac{dE_{cïm}}{dt}\,.
\end{equation}
For simplicity the same assumption as before has to be made, i.e. black holes exert unimportant kinetic pressure.
Thus, we get
\begin{eqnarray}\label{eqrhoaccr}
  \rho_{BH}&=&\frac{1}{a^3}\int ^{\infty} _{m_{ca}}A\,\zeta^{n-1}\,
  \exp\Big((n-1)\int ^{t} _{0}\rho_{rad} \,dt\Big) \,m^{-n+1}\,dm \nonumber\\
  &=&\frac{A}{a^3}\,\frac{1}{n-2}\,\zeta\,(k\,m_5)^{-n+2}\,\exp\Big(\int ^{t} _{0}\rho_{rad} \,dt\Big)\,.
\end{eqnarray}
We observe that in order to proceed further and be able to find the power of the time evolution of the black hole
density we have to know the integral $\exp(\int ^{t} _{0} \rho_{rad} \,dt)$ since $\rho_{BH}$ is proportional to it.
The comoving transfer rate per volume $\frac{dE_{cm}}{dt}$ is given by
\begin{eqnarray}\label{decdtaccr}
  \frac{dE_{cïm}}{dt}&=&\frac{A}{a^3}\,\frac{n-1}{n-2}\,\zeta \,k^{-n+2}\,m_5^{-n+2}\,\rho_{rad}\,
  \exp\Big(\int ^{t} _{0} \rho_{rad}\,dt\Big) \nonumber\\
  &&-\frac{A}{a^3}\,\zeta\,k^{-n+2}\,m_5^{-n+2}\,\rho_{rad}\,
  \exp\Big(\int ^{t} _{0} \rho_{rad}\,dt\Big) \,\Theta(m-m_{ca})\,.
\end{eqnarray}
The complete set of equations Eqs. (\ref{conservationaccr}), (\ref{expansionaccr}), (\ref{eqrhoaccr}) and
(\ref{decdtaccr}) form an integro-differential system and can be solved only numerically for various ranges of the
parameters.

\section{Conclusions}

The present study shows that the proposed baryogenesis scenario of accreting primordial black holes in a RS braneworld
is capable to generate efficient baryogenesis even for very small CP violating angles. In summary the key points are
\begin{itemize}
  \item The allowed by the mechanism BH mass range includes a mass spectrum around the higher dimensional Planck mass.
  The latter is important since this mass spectrum is energetically favorable to be generated from high energy
  interactions in the very early braneworld cosmic history.
  \item The baryogenesis process in a 5-dim RS cosmology becomes easier than in the standard 4-dim universe because of the
  accretion in the high energy regime.
  \item The Higgs sector has not to be necessarily that of the two-Higgs model. It just requires a Higgs sector with
  very low CP asymmetry.
  \item  It is not necessary the universe to be BH dominated at the time of the BHs creation since it is possible to
  turn into BH domination due to the accretion. However, since the proposed mechanism is able to generate very large
  baryon asymmetry, the black hole domination requirement is not crucial.
  \item The key point of producing large baryon asymmetry is the existence of an early high energy regime with an
  unconventional expansion rate that favors accretion. Thus, any alternative cosmological model bearing this feature
  can also give efficient baryogenesis.
\end{itemize}

\section{Acknowledgements}

We would like to acknowledge enlightening discussion with E.V. Bugaev, Anne Green, Hiroyuki Tashiro, Kazunori Kohri,
Jun'ichi Yokoyama and Raf Guedens.

\end{document}